\documentclass[%
reprint,
superscriptaddress,nobibnotes,
aps,prl,
 amsmath,amssymb,
 floatfix,
]{revtex4-1}
\usepackage{siunitx}
\usepackage[utf8]{inputenc}
\usepackage{graphicx,color}
\usepackage{dcolumn}
\usepackage{bm}
\usepackage{soul}

\begin{document}
\title{Activation of nominally silent domain wall-localized phonons from GHz to THz}

\author{Peng Chen}
\affiliation{Quantum Materials Theory, Istituto Italiano di Tecnologia,16163 Genova, Italy.}
\author{Louis Ponet}
\affiliation{Quantum Materials Theory, Istituto Italiano di Tecnologia,16163 Genova, Italy.}
\affiliation{Scuola Normale Superiore di Pisa, 56126 Pisa, Italy}
\author{Keji Lai}
\affiliation{Department of Physics, University of Texas at Austin, Austin, TX 78712, USA.}
\author{Roberto Cingolani}
\affiliation{Quantum Materials Theory, Istituto Italiano di Tecnologia,16163 Genova, Italy.}
\author{Sergey Artyukhin}
\affiliation{Quantum Materials Theory, Istituto Italiano di Tecnologia,16163 Genova, Italy.}
\pacs{Valid PACS appear here}
\maketitle
{\bf
Ferroelectric domain walls (DWs) are nanoscale topological defects that can be easily tailored to create nanoscale devices \cite{Scott2007,Tagantsev2010}. 
Their excitations, recently discovered to be responsible for DW GHz conductivity, hold promise for  faster signal transmission and processing speed compared to the existing technology \cite{Lee2003,Scott2012,Wu2017}. 
Here we find that DW phonons disperse from GHz to THz frequencies, thus explaining the origin of the surprisingly broad GHz signature in DW conductivity \cite{Wu2017}. Puzzling activation of nominally silent DW sliding modes in BiFeO$_3$ \cite{Huang2019} is traced back to DW tilting and resulting asymmetry in wall-localized phonons.
The obtained phonon spectra and selection rules are used to simulate scanning impedance microscopy, emerging as a powerful probe in nanophononics.
The results will guide experimental discovery of the predicted phonon branches and design of DW-based nanodevices.}

\begin{figure}[b]
   \includegraphics[width=0.99\linewidth]{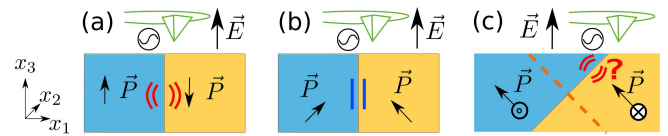}
 \caption{\label{fig:fig1}Illustrations of DW sliding modes interacting with the electric field of the SMIM tip: (a) excitation of a DW sliding mode is allowed when the field favors one of the domains; (b,c) when the field from the tip has the same projection on the polarizations of two domains, the excitation of the DW sliding mode is forbidden. (c) 71$^\circ$ domain wall corresponding to the $P_2$ reversal in BFO. The wall is tilted to minimize the elastic energy due to strain mismatch between the two domains, thus breaking the mirror indicated with a dashed line. The pseudocubic coordinates $x_1,x_2,x_3$ shown here are used throughout the paper and are referred to as the horizontal, out-of-plane and vertical directions.
 }
 \end{figure}

The seminal work of Seidel et al. \cite{Seidel2009} demonstrated that DC conductivity is higher at DWs in BiFeO$_3$ (BFO) than in the bulk, enabling signal transmission along the walls. 
This inspired a new paradigm for the design of DW-based nanoelectronic devices \cite{Catalan2012,Matsubara2015,McQuaid2017,Sharma2017,Huang2017,Mundy2017,Turner2018,Akamatsu2018,Schaab2018}, leading to demonstration of DW-based diode and transistor operating in kHz range \cite{Schaab2018,McQuaid2017}. The push for higher frequencies, used in modern computers, has led to a discovery of giant increase of the effective DW conductivity at GHz frequencies in recent microwave microscopy experiments  \cite{Tselev2016,Wu2017,Prosandeev2018}. The conductivity results from excitation of soft DW-localized phonon modes, that are at the heart of switching, microwave dielectric loss and dielectric constant enhancement in ferroelectrics  \cite{Wu2017,Hlinka2017}. They correspond to oscillations of the DW plane and can be excited by an AC electric field that favors one of the domains during its half period (Fig.~\ref{fig:fig1}(a)). Here we show that the frequency of these phonons can go from GHz all the way up to THz, going beyond the frequency range of modern surface acoustic wave-based cell phone transducers.

The results explain why the frequency range of the SMIM anomaly is so wide and reveal DW-localized bands in the phonon spectra of ferroelectrics. 
We also address the peculiar selection rules for DW excitation. When the ferroelectric polarization component along the driving field is the same in the two domains across the wall, the DW vibration should not be excited (Fig.~\ref{fig:fig1}(b)), as seen at 180$^\circ$ walls on the $ac$ face of YMnO$_3$ \cite{Wu2017}. 
However, this rule is violated in recent experiments on  rhombohedral BFO, where polarization components along the field across the wall do not differ (Fig.~\ref{fig:fig1}(c)), while the DW mode is still excited \cite{Huang2019}. 

The results could inspire the exploration of DW-based nanosystems in THz frequency range, and lead to novel DW-based phononic nanodevices.

\begin{figure}[!t]
 \includegraphics*[width=0.99\linewidth]{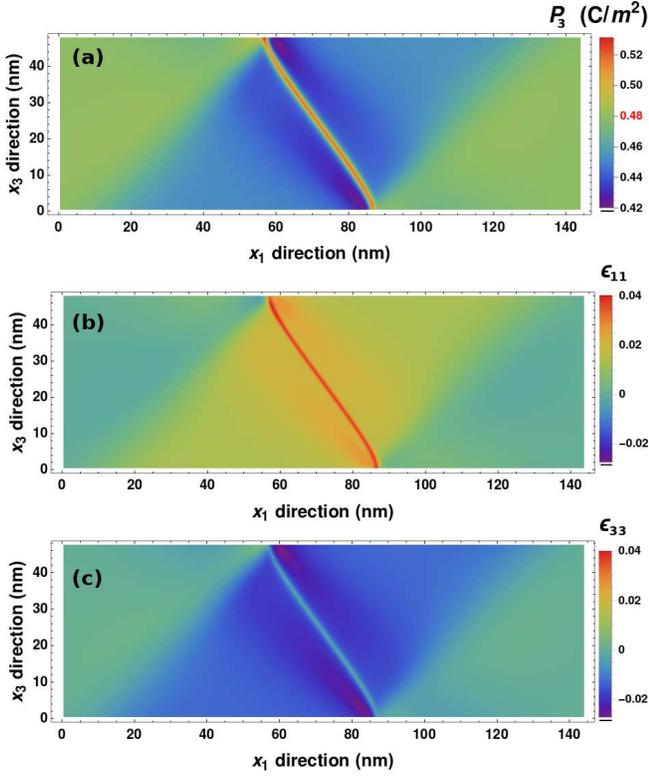}
 \caption{
 \label{fig:Strain}
Asymmetric polarization (a) and strain profiles (b,c) at the domain wall obtained by numerical energy minimization. Each polarization component in the bulk takes the value, marked in red on the color scale in (a).}
\end{figure}\begin{figure*}[!ht]
 \includegraphics*[width=\linewidth]{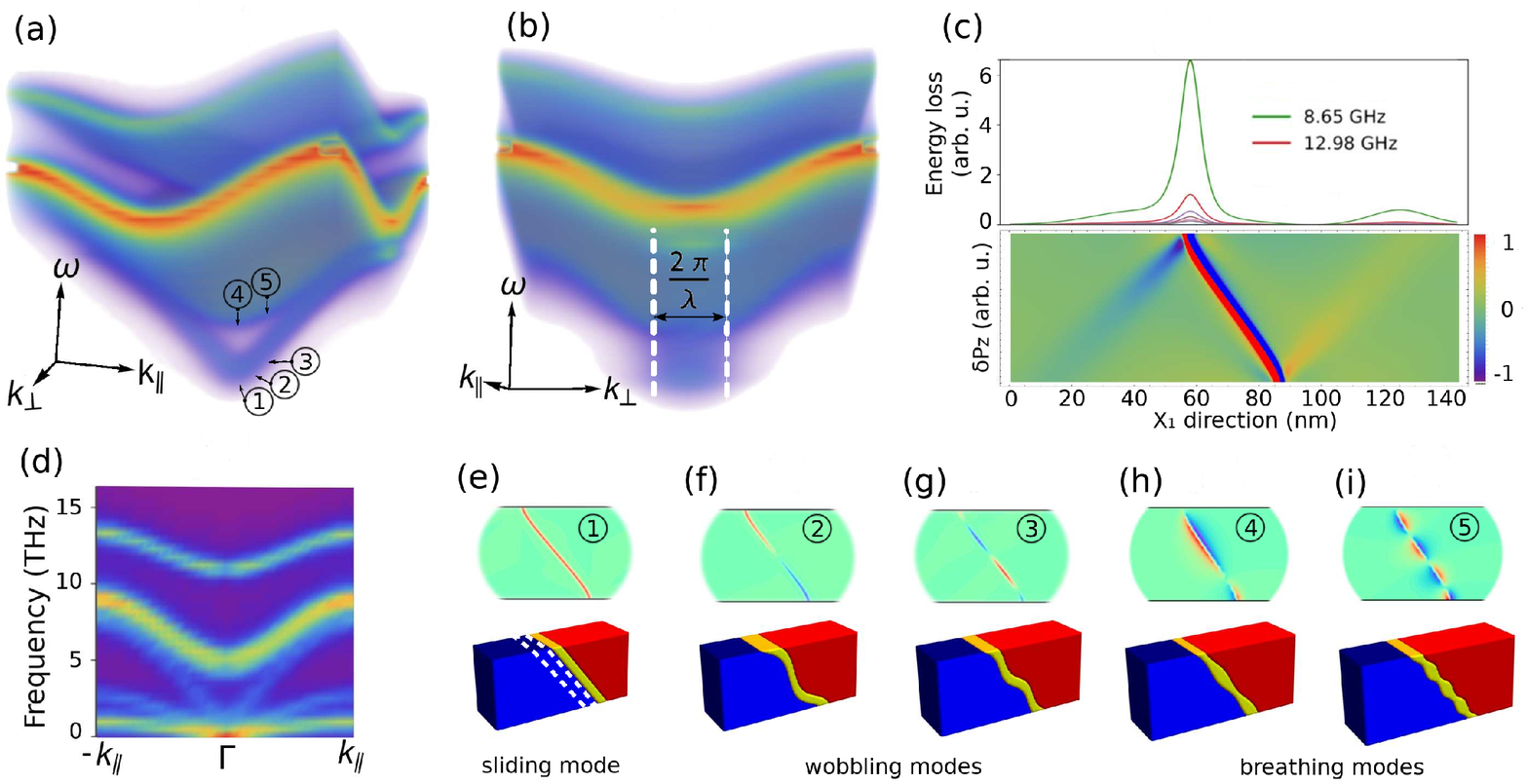}
 \vskip -0.5cm
 \caption{\label{fig:modes}
(a,b) The phonon spectral function for polar modes in BFO with an R71 DW. The intense bands are the polar modes. The faint stripe at low energy, merging with the phonon band away from the zone center corresponds to the DW sliding modes. The labels in (a), \textcircled{1}$\sim$\textcircled{5}, are used to indicate the DW localized modes in (e,f,g,h,i) with correspondingly the same labels. The spectral weight extends to $k\sim 2\pi/\lambda$ in the direction perpendicular to the wall, as seen in panel (b). (c) The simulated SMIM loss signal across R71 DW $I(r,\omega)$. A peak at the DW and a signature from the long-range strain profile are seen at the sliding mode frequency. The notable asymmetry in the polarization profile, $\delta P_3 (r)$ for the DW sliding mode that leads to the excitation of this mode by the vertical electric field is shown below; (d) the spectral function for both polar and acoustic phonons along the $k_{\parallel}$ direction (a cross section plot of (a) and (b) through $\gamma$ point).
 (e-i) Real-space polarization profile $\delta P_2(r)$ of the low frequency DW sliding (e-g) and breathing (h,i) modes (upper row), 
 along with schematic illustrations of corresponding DW vibrations (below). The initial position of the DW is marked with white dashed lines.}
\end{figure*}

BFO is a rare room-temperature multiferroic with the largest spontaneous polarization among single-phase compounds known to date, and may be soon used in nano-devices \cite{Ramesh2018}.   Three types of ferroelectric DWs in the rhomohedral phase are distinguished by the number of polarization components being reversed \cite{Marton2010}. The most puzzling data is on R71$^\circ$ DWs, at which the $P_2$ component is reversed, and the polarization rotates by \ang{71} across the wall, as seen in Fig.~\ref{fig:fig1}(c). R71 DWs are often seen in BFO films grown on a $[001]$ oriented substrate \cite{Ziegler2013,Domingo2017}, such as SrTiO$_3$ and DyScO$_3$. The sliding mode of this wall should not be excited by the field normal to the (100) surface, since neither domain is favored by the field, as seen in Fig.~\ref{fig:fig1}(c). However, the impedance signal is still observed experimentally \cite{Huang2019}. The important distinction between \ang{180} DWs in YMnO$_3$ and R71 DWs in BFO is their orientation with respect to the surface. YMnO$_3$ DWs are mostly normal to the surface, while BFO R71 DWs are tilted away from the surface normal. In that case the combination of the surface and DW orientations breaks the symmetry, as seen in Fig.~\ref{fig:fig1}(c).
This symmetry breaking plays a key role in their response to electric fields, even when the simple selection rule, summarised in Fig.~\ref{fig:fig1}(a,b), suggests no response. 


{\bf Results}


\begin{figure}[t]
 \includegraphics*[width=\linewidth]{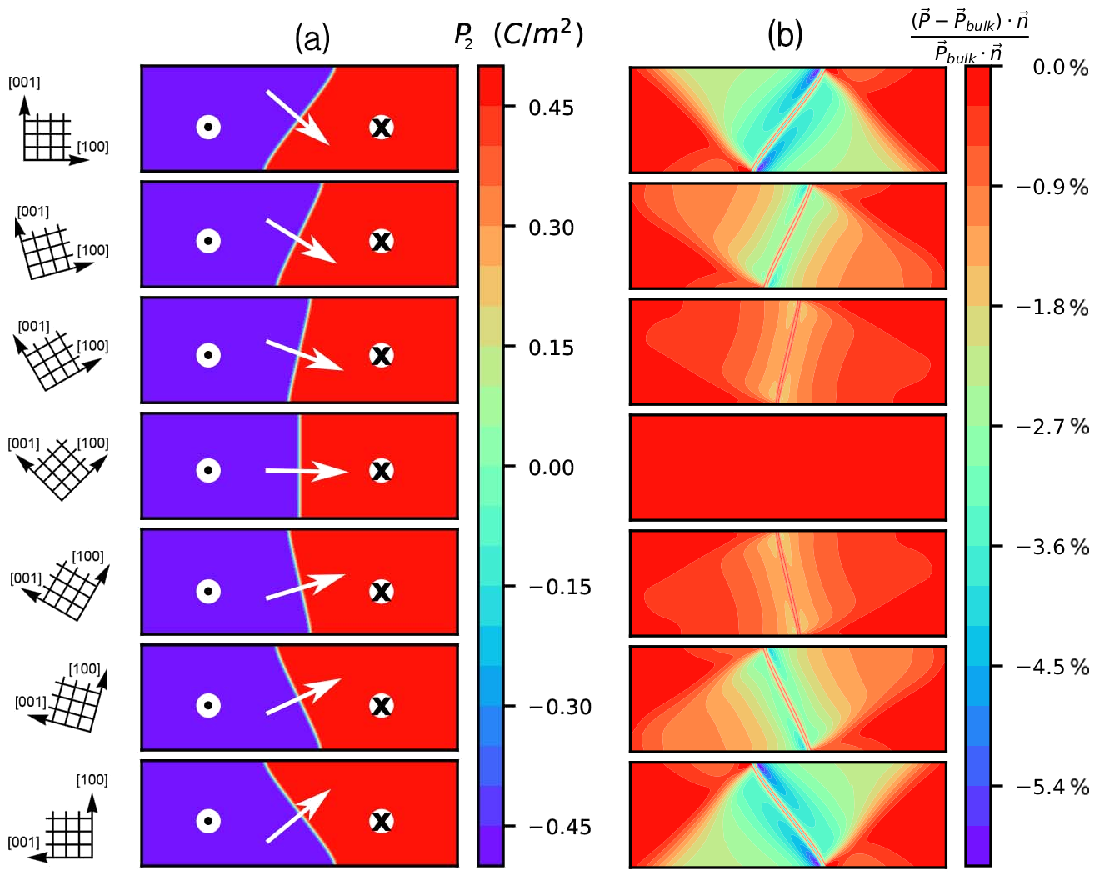}
 \caption{
 \label{fig:PzOrientation}
 (a) Thin films with various surface orientations with respect to the crystallographic directions, indicated on the left. The DW normal is indicated by the arrow. (b) Normalized deviation of the vertical component of polarization $(P_3(r)-P_{3,bulk}/P_{3,bulk}$ from its bulk value $P_{3,bulk}$ due to the stress at the DW.}
\end{figure}

We use the Ginzburg-Landau-Devonshire (GLD) model for DWs in BFO to capture the energetics of interacting ferroelectric polarization and strain (see Methods for details). The DW is tilted 45$^\circ$ away from the surface normal and bends near the surface to reduce the DW area, as shown in Fig.~\ref{fig:Strain}. Long-range strain textures emanate from the bent segments. $P_3$ decrease in the blue area near the surface is driven by the compression $\epsilon_{33}$ and the tensile strain $\epsilon_{11}$ via electrostriction.

The vibrational modes of the system with DWs are computed from the linearized equations of motion within a discretized Ginzburg-Landau model, and shown in Fig.~\ref{fig:modes}(a,b) (see the Methods section for details). The DW breaks translational invariance and therefore the modes are not characterized by a well-defined quasimomentum. The traditional phonon band structure is then replaced by a spectral function, as seen in Fig.~\ref{fig:modes}(a), where the intensity at frequency $\omega$ and wavevector $k$ indicates the content of plane waves with that wavevector in the eigenmodes at that frequency. For a translationally invariant system, sharp peaks would then appear in the spectral function ${A(\omega,\vec k)=\sum_\nu \delta(\omega\pm\omega_{\nu})|\langle\delta P_{2,\nu}|\exp(i\vec k\cdot\vec r)\rangle|^2}$.
The long-ranged strain texture of the wall, shown in Fig.~\ref{fig:Strain}, is responsible for strong mixing between the DW sliding modes and acoustic phonons.

Fig.~\ref{fig:modes}(a,b) shows the content of plane waves with $P_2\sim e^{i\vec k\vec r}$, while Fig.~\ref{fig:modes}(d) shows the phonon dispersion along the wall. At low frequencies acoustic phonons are observed (their intensity is divided by 10 in Fig.~\ref{fig:modes}(d) to make the DW branch visible). They correspond to strain modulations and mix with $P_2$ modes due to electrostriction,
${f_{q}=-\frac{1}{2}\epsilon_{ij}q_{ijkl}P_{k}P_{l}\label{eq:qpp1}}$. The V-shaped low frequency branch, marked with (1-3) in Fig.~\ref{fig:modes}(a), extends from around 10~GHz at $\Gamma$-point all the way to the bulk polar phonons, and corresponds to the DW sliding and wobbling modes, illustrated in  Fig.~\ref{fig:modes}(e-g). The phonons in this branch disperse along the wall (along $k_\parallel$), but are localized in the perpendicular direction, and therefore their Fourier components extend to $k_\perp\sim\pm\pi/\lambda$, where $\lambda$ stands for the DW width. The intensity of the DW-localized branch is lower than that of the bulk phonons due to the low volume fraction occupied by DWs. The higher frequency modes are the bulk polar phonons.

Fig.~\ref{fig:modes}(e-i) show $\delta P_2$ profiles corresponding to some of the lowest frequency DW sliding modes(e,f,g) and breathing modes (h,i). The nodeless mode shown in the upper panel of Fig.~\ref{fig:modes}(e) corresponds to the $P_2$ increase at the wall during half-period of the oscillation, therefore adding the DW area to the positive domain. During another half-period the negative domain grows. Therefore this mode can be thought of as DW sliding, depicted schematically in the lower panels of Fig.~\ref{fig:modes}(e). A higher energy mode with a node in that branch, shown in the upper panel of Fig.~\ref{fig:modes}(f), corresponds to DW wobbling.
The DW breathing modes, schematically shown in Fig.~\ref{fig:modes}(h,i), are found below the band of bulk polar modes, and are indicated with markers (4,5) in Fig.~\ref{fig:modes}(a).

Now we move to the origin of the SMIM signal. 
In the experiments an AC electric field is applied between the tip and the back electrode, as shown in Fig.~\ref{fig:fig1}, and the current is measured. 
The phonons that have a non-zero energy in the oscillating field of the tip, ${-\int dr \vec{\delta P}(\vec r)\cdot\vec E(\vec r)e^{i\omega t}}$, are excited and give rise to a displacement current component in phase with the field. The corresponding loss, ${\mathrm{Re} \int dr\vec j(r)\cdot\vec E(r)}$ is measured in addition to the Ohmic losses due to itinerant electrons. The amplitude $x_i$ of a phonon mode $i$ with an eigenfrequency $\omega_i$ is governed by the equation of motion
\begin{equation}
m_i \ddot{x}_i+\gamma \dot{x}_i+m_i\omega_i^2x_i=\int dr \delta\vec P_i(\vec r)\cdot\vec E(\vec r)e^{i\omega t},
\end{equation}
where the mode is characterized by its effective mass $m_i$, damping $\gamma_i$ and the spatial polarization profile $\delta P_i(r)$. The oscillating driving force on the right hand side is due to the electric field $\vec{E}(\vec{r})e^{i\omega t}$ of the SMIM tip. Looking for the solution in the form $x_i=x_{i0}e^{i\omega t}$, we obtain the loss power at the tip position $\vec r$,
\begin{equation}
I(\vec r,\omega)=\omega^2\gamma\sum_i\frac{\left[\int dr'\:\vec E(\vec{r'}-\vec r)\cdot\delta\vec P_i(\vec {r'})\right]^2}{m_i^2(\omega_i^2-\omega^2)^2+\omega^2\gamma^2}.
\end{equation}

The frequency dependence is characterized by a Lorentzian. The integral in the numerator is negligible for phonons whose spatial oscillation period is much smaller that the length scale of the electric field inhomogeneity, roughly determined by the tip radius $R$, therefore the phonons with wavevectors $k\lessapprox R^{-1}$ are excited. Fig.~\ref{fig:modes}(c) shows the simulated signal for a BFO sample containing a R71 DW. An asymmetric peak at the DW and a weak signature due to the long-range strain features are visible.

{\bf Discussion}

Figure~\ref{fig:modes}(c) shows that SMIM conductivity is much higher at DW than in the bulk due to low-frequency DW sliding and wobbling modes. Their dispersion spans the whole GHz range and extends to the bulk polar phonon band, usually positioned at THz frequency.  This explains why the GHz microwave conductivity at DWs in hexagonal manganites does not show a narrow peak in frequency domain, but rather rises monotonically towards higher frequencies. The proximity of DW phonons to acoustic branches also leads to phonon scattering and affects thermal conductivity \cite{Royo2017}. The low frequency of sliding modes is also behind the giant enhancement of the static dielectric permittivity \cite{Hlinka2017}. In addition, soft DW phonons are responsible for mechanical softening phonomena, recently observed in several ferroelectrics \cite{Stefani2025}.

In materials with strong strain-polarization interactions, DWs are aligned into regular lattices by strain fields \cite{Artyukhin2013,Wang2014}, and spectral features similar to the ones in Fig.~\ref{fig:modes}(a,b) are expected. Alternatively, when DWs are randomly oriented, the spectral weigth of DW wobbling and breathing modes spreads along different directions, but the phonon density of states is expected to retain its shape, with the peak at GHz frequencies. The directionality of DWs
and their populations in the sample can then be inferred from the orientation and intensity of low-energy branches, as seen in Fig.~S2.

The recent observation of nominally silent DW modes in BFO \cite{Huang2019} being activated is a surprising evidence that DW type and orientation control local AC conductivity. Possible DW orientations have been classified using symmetry and compatibility analyses \cite{Janovec1976,Fousek1969}. In essence, the distances between ionic planes in the two domains must match along the wall. This is only possible for particular DW  orientations, for which the strain components in the DW plane match at the wall (see Eq.~S3). The energy penalty for violating this condition scales linearly with the wall area, therefore fixing the DW orientation in the bulk. However, in thin films surface strain relaxation allows the DW to deviate from its optimal bulk orientation.

In an infinite bulk sample with a DW, a diagonal mirror plane $M_{dw}(-1, 0, 1)$, shown by the dashed line in Fig.~\ref{fig:fig1}(c), is a symmetry operation. This symmetry imposes the requirements: $ \epsilon_{11}=\epsilon_{33}, \epsilon_{12}=\epsilon_{23}$. 
The surface of the thin film and the domain wall orientation together break this mirror symmetry $m_{dw}$ if they are not orthogonal.
The stress at the tilted domain wall $\sigma_{ij}=\partial f/\partial \epsilon_{ij}$ then acts on geometrically different regions from the left and right sides of the wall. The resulting asymmetric strain induces asymmetric polarization across the wall via electrostriction Eq.~\ref{eq:qpp}. This phenomenon is observed in Fig.~\ref{fig:PzOrientation}, where we simulated thin films with different surface orientations with respect to the crystallographic directions (different ways of cutting a film out of a slab). It is seen that the DW orients perpendicular to the $[P_1,0,P_3]$ direction, and a small bending near the surface is observed in Fig.~\ref{fig:PzOrientation}(a) \cite{Ishibashi2002,Yoshihiro2005,Conti2004,Linze2018}. In very thin films, the wall deviates from the $[101]$ orientation towards the surface normal. Fig.~\ref{fig:PzOrientation}(b) shows the polarization component normal to the surface. When the wall is tilted at 45$^\circ$, the asymmetry of the polarization $P_3$ is evident in Fig.~\ref{fig:Strain}(a), especially near the surface where the polarization deviates the most from the bulk value. This asymmetry comes from the  asymmetric strain, plotted in Fig.~\ref{fig:Strain}(b,c).
Even though the DW polarization profile is very narrow, the accompanying strain texture, emanating from the areas of DW bending near the surface \cite{Ishibashi2002,Yoshihiro2005,Conti2004,Linze2018}, extends surprisingly far \cite{Salje1993,Salje2016}, as seen in Fig.~\ref{fig:Strain}(b,c).

The polarization asymmetry eventually results in the asymmetry in the phonon polarization profile of the DW sliding mode, as seen in the lower panel of Fig.~\ref{fig:modes}(c). When the polarization $P_3$ interacts with the external electric field, normal to the surface, the electrostatic energy $-\int dr\vec{E(r)}\cdot\vec{P}(r-r_{DW})$ results in a force on the wall due to $\vec{E}\cdot\frac{\partial\vec{P}(r-r_{DW})}{\partial x_{1,DW}}\neq0$ and excitation of the DW vibration by the SMIM field.

In addition to SMIM, it may be possible to observe the DW-localized modes with inelastic neutron or X-ray scattering, although their intensity may be low due to low volume fraction occupied by DWs. Rapid quenching of a sample across the ferroelectric transition, leading to high DW densities \cite{Griffin12}, or strain-induced generation and alignment of ferroelastic DWs \cite{Chae2012} should increase the intensity of these branches and enable their observation in inelastic neutron scattering experiments and other bulk spectroscopies.

In summary, the simulated phonon spectra reveal the DW-localized phonon branches starting from GHz and extending all the way to THz frequencies. This explains wide frequency range of the conductivity anomaly observed in hexagonal manganites \cite{Wu2017} and other materials. The surprising activation of the nominally silent DW mode at R71$^\circ$ DWs in BFO is interpreted in terms of phonon polarization asymmetry due to the interplay of electrostriction and elastic compatibility at a tilted DW. 
The proposed way to simulate SMIM experiments may be used in emerging second-principles methodologies.
Since the low-energy theory used here is rather general, similar phonon spectra must be expected for all ferroic materials, where the order parameter couples strongly to the lattice.
We hope this work will motivate the experimental search for DW-localized phonon branches, guide the interpretation of SMIM studies and eventually inspire the development of high-speed DW-based phononic devices, extending surface acoustic wave-based technology to THz DW-based circuits.

{\bf References}
\bibliography{R71DW}

\begin{thebibliography}{46}%
\makeatletter
\providecommand \@ifxundefined [1]{%
 \@ifx{#1\undefined}
}%
\providecommand \@ifnum [1]{%
 \ifnum #1\expandafter \@firstoftwo
 \else \expandafter \@secondoftwo
 \fi
}%
\providecommand \@ifx [1]{%
 \ifx #1\expandafter \@firstoftwo
 \else \expandafter \@secondoftwo
 \fi
}%
\providecommand \natexlab [1]{#1}%
\providecommand \enquote  [1]{``#1''}%
\providecommand \bibnamefont  [1]{#1}%
\providecommand \bibfnamefont [1]{#1}%
\providecommand \citenamefont [1]{#1}%
\providecommand \href@noop [0]{\@secondoftwo}%
\providecommand \href [0]{\begingroup \@sanitize@url \@href}%
\providecommand \@href[1]{\@@startlink{#1}\@@href}%
\providecommand \@@href[1]{\endgroup#1\@@endlink}%
\providecommand \@sanitize@url [0]{\catcode `\\12\catcode `\$12\catcode
  `\&12\catcode `\#12\catcode `\^12\catcode `\_12\catcode `\%12\relax}%
\providecommand \@@startlink[1]{}%
\providecommand \@@endlink[0]{}%
\providecommand \url  [0]{\begingroup\@sanitize@url \@url }%
\providecommand \@url [1]{\endgroup\@href {#1}{\urlprefix }}%
\providecommand \urlprefix  [0]{URL }%
\providecommand \Eprint [0]{\href }%
\providecommand \doibase [0]{http://dx.doi.org/}%
\providecommand \selectlanguage [0]{\@gobble}%
\providecommand \bibinfo  [0]{\@secondoftwo}%
\providecommand \bibfield  [0]{\@secondoftwo}%
\providecommand \translation [1]{[#1]}%
\providecommand \BibitemOpen [0]{}%
\providecommand \bibitemStop [0]{}%
\providecommand \bibitemNoStop [0]{.\EOS\space}%
\providecommand \EOS [0]{\spacefactor3000\relax}%
\providecommand \BibitemShut  [1]{\csname bibitem#1\endcsname}%
\let\auto@bib@innerbib\@empty
\bibitem [{\citenamefont {Scott}(2007)}]{Scott2007}%
  \BibitemOpen
  \bibfield  {author} {\bibinfo {author} {\bibfnamefont {J.~F.}\ \bibnamefont
  {Scott}},\ }\href@noop {} {\bibfield  {journal} {\bibinfo  {journal}
  {Science}\ }\textbf {\bibinfo {volume} {315}},\ \bibinfo {pages} {954}
  (\bibinfo {year} {2007})}\BibitemShut {NoStop}%
\bibitem [{\citenamefont {Tagantsev}\ \emph {et~al.}(2010)\citenamefont
  {Tagantsev}, \citenamefont {Cross},\ and\ \citenamefont
  {Fousek}}]{Tagantsev2010}%
  \BibitemOpen
  \bibfield  {author} {\bibinfo {author} {\bibfnamefont {A.~K.}\ \bibnamefont
  {Tagantsev}}, \bibinfo {author} {\bibfnamefont {L.~E.}\ \bibnamefont
  {Cross}}, \ and\ \bibinfo {author} {\bibfnamefont {J.}~\bibnamefont
  {Fousek}},\ }\href@noop {} {\emph {\bibinfo {title} {{Domains in Ferroic
  Crystals and Thin Films}}}}\ (\bibinfo  {publisher} {Springer New York},\
  \bibinfo {address} {New York, NY},\ \bibinfo {year} {2010})\BibitemShut
  {NoStop}%
\bibitem [{\citenamefont {Lee}\ \emph {et~al.}(2003)\citenamefont {Lee},
  \citenamefont {Salje},\ and\ \citenamefont {Bismayer}}]{Lee2003}%
  \BibitemOpen
  \bibfield  {author} {\bibinfo {author} {\bibfnamefont {W.~T.}\ \bibnamefont
  {Lee}}, \bibinfo {author} {\bibfnamefont {E.~K.~H.}\ \bibnamefont {Salje}}, \
  and\ \bibinfo {author} {\bibfnamefont {U.}~\bibnamefont {Bismayer}},\
  }\href@noop {} {\bibfield  {journal} {\bibinfo  {journal} {Journal of
  Physics: Condensed Matter}\ }\textbf {\bibinfo {volume} {15}},\ \bibinfo
  {pages} {1353} (\bibinfo {year} {2003})}\BibitemShut {NoStop}%
\bibitem [{\citenamefont {Scott}\ \emph {et~al.}(2012)\citenamefont {Scott},
  \citenamefont {Salje},\ and\ \citenamefont {Carpenter}}]{Scott2012}%
  \BibitemOpen
  \bibfield  {author} {\bibinfo {author} {\bibfnamefont {J.~F.}\ \bibnamefont
  {Scott}}, \bibinfo {author} {\bibfnamefont {E.~K.~H.}\ \bibnamefont {Salje}},
  \ and\ \bibinfo {author} {\bibfnamefont {M.~A.}\ \bibnamefont {Carpenter}},\
  }\href@noop {} {\bibfield  {journal} {\bibinfo  {journal} {Phys. Rev. Lett.}\
  }\textbf {\bibinfo {volume} {109}},\ \bibinfo {pages} {187601} (\bibinfo
  {year} {2012})}\BibitemShut {NoStop}%
\bibitem [{\citenamefont {Wu}\ \emph {et~al.}(2017)\citenamefont {Wu},
  \citenamefont {Petralanda}, \citenamefont {Zheng}, \citenamefont {Ren},
  \citenamefont {Hu}, \citenamefont {Cheong}, \citenamefont {Artyukhin},\ and\
  \citenamefont {Lai}}]{Wu2017}%
  \BibitemOpen
  \bibfield  {author} {\bibinfo {author} {\bibfnamefont {X.}~\bibnamefont
  {Wu}}, \bibinfo {author} {\bibfnamefont {U.}~\bibnamefont {Petralanda}},
  \bibinfo {author} {\bibfnamefont {L.}~\bibnamefont {Zheng}}, \bibinfo
  {author} {\bibfnamefont {Y.}~\bibnamefont {Ren}}, \bibinfo {author}
  {\bibfnamefont {R.}~\bibnamefont {Hu}}, \bibinfo {author} {\bibfnamefont
  {S.-W.}\ \bibnamefont {Cheong}}, \bibinfo {author} {\bibfnamefont
  {S.}~\bibnamefont {Artyukhin}}, \ and\ \bibinfo {author} {\bibfnamefont
  {K.}~\bibnamefont {Lai}},\ }\href@noop {} {\bibfield  {journal} {\bibinfo
  {journal} {Sci. Adv.}\ }\textbf {\bibinfo {volume} {3}},\ \bibinfo {pages}
  {e1602371} (\bibinfo {year} {2017})}\BibitemShut {NoStop}%
\bibitem [{\citenamefont {Huang}\ \emph {et~al.}()\citenamefont {Huang},
  \citenamefont {Zheng}, \citenamefont {Chen}, \citenamefont {Cheng},
  \citenamefont {Ponet}, \citenamefont {Ramesh}, \citenamefont {Chen},
  \citenamefont {Artyukhin}, \citenamefont {Chu},\ and\ \citenamefont
  {Lai}}]{Huang2019}%
  \BibitemOpen
  \bibfield  {author} {\bibinfo {author} {\bibfnamefont {Y.~L.}\ \bibnamefont
  {Huang}}, \bibinfo {author} {\bibfnamefont {L.}~\bibnamefont {Zheng}},
  \bibinfo {author} {\bibfnamefont {P.}~\bibnamefont {Chen}}, \bibinfo {author}
  {\bibfnamefont {X.}~\bibnamefont {Cheng}}, \bibinfo {author} {\bibfnamefont
  {L.}~\bibnamefont {Ponet}}, \bibinfo {author} {\bibfnamefont
  {R.}~\bibnamefont {Ramesh}}, \bibinfo {author} {\bibfnamefont {L.~Q.}\
  \bibnamefont {Chen}}, \bibinfo {author} {\bibfnamefont {S.}~\bibnamefont
  {Artyukhin}}, \bibinfo {author} {\bibfnamefont {Y.-H.}\ \bibnamefont {Chu}},
  \ and\ \bibinfo {author} {\bibfnamefont {K.}~\bibnamefont {Lai}},\
  }\href@noop {} {\bibinfo  {journal} {in preparation}\ }\BibitemShut {NoStop}%
\bibitem [{\citenamefont {Seidel}\ \emph {et~al.}(2009)\citenamefont {Seidel},
  \citenamefont {Martin}, \citenamefont {He}, \citenamefont {Zhan},
  \citenamefont {Chu}, \citenamefont {Rother}, \citenamefont {Hawkridge},
  \citenamefont {Maksymovych}, \citenamefont {Yu}, \citenamefont {Gajek},
  \citenamefont {Balke}, \citenamefont {Kalinin}, \citenamefont {Gemming},
  \citenamefont {Wang}, \citenamefont {Catalan}, \citenamefont {Scott},
  \citenamefont {Spaldin}, \citenamefont {Orenstein},\ and\ \citenamefont
  {Ramesh}}]{Seidel2009}%
  \BibitemOpen
\bibfield  {journal} {  }\bibfield  {author} {\bibinfo {author} {\bibfnamefont
  {J.}~\bibnamefont {Seidel}}, \bibinfo {author} {\bibfnamefont {L.~W.}\
  \bibnamefont {Martin}}, \bibinfo {author} {\bibfnamefont {Q.}~\bibnamefont
  {He}}, \bibinfo {author} {\bibfnamefont {Q.}~\bibnamefont {Zhan}}, \bibinfo
  {author} {\bibfnamefont {Y.-H.}\ \bibnamefont {Chu}}, \bibinfo {author}
  {\bibfnamefont {A.}~\bibnamefont {Rother}}, \bibinfo {author} {\bibfnamefont
  {M.~E.}\ \bibnamefont {Hawkridge}}, \bibinfo {author} {\bibfnamefont
  {P.}~\bibnamefont {Maksymovych}}, \bibinfo {author} {\bibfnamefont
  {P.}~\bibnamefont {Yu}}, \bibinfo {author} {\bibfnamefont {M.}~\bibnamefont
  {Gajek}}, \bibinfo {author} {\bibfnamefont {N.}~\bibnamefont {Balke}},
  \bibinfo {author} {\bibfnamefont {S.~V.}\ \bibnamefont {Kalinin}}, \bibinfo
  {author} {\bibfnamefont {S.}~\bibnamefont {Gemming}}, \bibinfo {author}
  {\bibfnamefont {F.}~\bibnamefont {Wang}}, \bibinfo {author} {\bibfnamefont
  {G.}~\bibnamefont {Catalan}}, \bibinfo {author} {\bibfnamefont {J.~F.}\
  \bibnamefont {Scott}}, \bibinfo {author} {\bibfnamefont {N.~A.}\ \bibnamefont
  {Spaldin}}, \bibinfo {author} {\bibfnamefont {J.}~\bibnamefont {Orenstein}},
  \ and\ \bibinfo {author} {\bibfnamefont {R.}~\bibnamefont {Ramesh}},\
  }\href@noop {} {\bibfield  {journal} {\bibinfo  {journal} {Nature Materials}\
  }\textbf {\bibinfo {volume} {8}},\ \bibinfo {pages} {229} (\bibinfo {year}
  {2009})}\BibitemShut {NoStop}%
\bibitem [{\citenamefont {Catalan}\ \emph {et~al.}(2012)\citenamefont
  {Catalan}, \citenamefont {Seidel}, \citenamefont {Ramesh},\ and\
  \citenamefont {Scott}}]{Catalan2012}%
  \BibitemOpen
  \bibfield  {author} {\bibinfo {author} {\bibfnamefont {G.}~\bibnamefont
  {Catalan}}, \bibinfo {author} {\bibfnamefont {J.}~\bibnamefont {Seidel}},
  \bibinfo {author} {\bibfnamefont {R.}~\bibnamefont {Ramesh}}, \ and\ \bibinfo
  {author} {\bibfnamefont {J.~F.}\ \bibnamefont {Scott}},\ }\href@noop {}
  {\bibfield  {journal} {\bibinfo  {journal} {Rev Mod Phys}\ }\textbf {\bibinfo
  {volume} {84}},\ \bibinfo {pages} {119} (\bibinfo {year} {2012})}\BibitemShut
  {NoStop}%
\bibitem [{\citenamefont {Matsubara}\ \emph {et~al.}(2015)\citenamefont
  {Matsubara}, \citenamefont {Manz}, \citenamefont {Mochizuki}, \citenamefont
  {Kubacka}, \citenamefont {Iyama}, \citenamefont {Aliouane}, \citenamefont
  {Kimura}, \citenamefont {Johnson}, \citenamefont {Meier},\ and\ \citenamefont
  {Fiebig}}]{Matsubara2015}%
  \BibitemOpen
  \bibfield  {author} {\bibinfo {author} {\bibfnamefont {M.}~\bibnamefont
  {Matsubara}}, \bibinfo {author} {\bibfnamefont {S.}~\bibnamefont {Manz}},
  \bibinfo {author} {\bibfnamefont {M.}~\bibnamefont {Mochizuki}}, \bibinfo
  {author} {\bibfnamefont {T.}~\bibnamefont {Kubacka}}, \bibinfo {author}
  {\bibfnamefont {A.}~\bibnamefont {Iyama}}, \bibinfo {author} {\bibfnamefont
  {N.}~\bibnamefont {Aliouane}}, \bibinfo {author} {\bibfnamefont
  {T.}~\bibnamefont {Kimura}}, \bibinfo {author} {\bibfnamefont {S.~L.}\
  \bibnamefont {Johnson}}, \bibinfo {author} {\bibfnamefont {D.}~\bibnamefont
  {Meier}}, \ and\ \bibinfo {author} {\bibfnamefont {M.}~\bibnamefont
  {Fiebig}},\ }\href@noop {} {\bibfield  {journal} {\bibinfo  {journal}
  {Science}\ }\textbf {\bibinfo {volume} {348}},\ \bibinfo {pages} {1112}
  (\bibinfo {year} {2015})}\BibitemShut {NoStop}%
\bibitem [{\citenamefont {McQuaid}\ \emph {et~al.}(2017)\citenamefont
  {McQuaid}, \citenamefont {Campbell}, \citenamefont {Whatmore}, \citenamefont
  {Kumar},\ and\ \citenamefont {Gregg}}]{McQuaid2017}%
  \BibitemOpen
  \bibfield  {author} {\bibinfo {author} {\bibfnamefont {R.~G.~P.}\
  \bibnamefont {McQuaid}}, \bibinfo {author} {\bibfnamefont {M.~P.}\
  \bibnamefont {Campbell}}, \bibinfo {author} {\bibfnamefont {R.~W.}\
  \bibnamefont {Whatmore}}, \bibinfo {author} {\bibfnamefont {A.}~\bibnamefont
  {Kumar}}, \ and\ \bibinfo {author} {\bibfnamefont {J.~M.}\ \bibnamefont
  {Gregg}},\ }\href@noop {} {\bibfield  {journal} {\bibinfo  {journal} {Nature
  Communications}\ }\textbf {\bibinfo {volume} {8}},\ \bibinfo {pages} {15105}
  (\bibinfo {year} {2017})}\BibitemShut {NoStop}%
\bibitem [{\citenamefont {Sharma}\ \emph {et~al.}(2017)\citenamefont {Sharma},
  \citenamefont {Zhang}, \citenamefont {Sando}, \citenamefont {Lei},
  \citenamefont {Liu}, \citenamefont {Li}, \citenamefont {Nagarajan},\ and\
  \citenamefont {Seidel}}]{Sharma2017}%
  \BibitemOpen
  \bibfield  {author} {\bibinfo {author} {\bibfnamefont {P.}~\bibnamefont
  {Sharma}}, \bibinfo {author} {\bibfnamefont {Q.}~\bibnamefont {Zhang}},
  \bibinfo {author} {\bibfnamefont {D.}~\bibnamefont {Sando}}, \bibinfo
  {author} {\bibfnamefont {C.~H.}\ \bibnamefont {Lei}}, \bibinfo {author}
  {\bibfnamefont {Y.}~\bibnamefont {Liu}}, \bibinfo {author} {\bibfnamefont
  {J.}~\bibnamefont {Li}}, \bibinfo {author} {\bibfnamefont {V.}~\bibnamefont
  {Nagarajan}}, \ and\ \bibinfo {author} {\bibfnamefont {J.}~\bibnamefont
  {Seidel}},\ }\href@noop {} {\bibfield  {journal} {\bibinfo  {journal} {Sci.
  Adv.}\ }\textbf {\bibinfo {volume} {3}},\ \bibinfo {pages} {e1700512}
  (\bibinfo {year} {2017})}\BibitemShut {NoStop}%
\bibitem [{\citenamefont {Huang}\ and\ \citenamefont
  {Cheong}(2017)}]{Huang2017}%
  \BibitemOpen
  \bibfield  {author} {\bibinfo {author} {\bibfnamefont {F.-T.}\ \bibnamefont
  {Huang}}\ and\ \bibinfo {author} {\bibfnamefont {S.-W.}\ \bibnamefont
  {Cheong}},\ }\href@noop {} {\bibfield  {journal} {\bibinfo  {journal} {Nature
  Reviews Materials}\ }\textbf {\bibinfo {volume} {2}},\ \bibinfo {pages}
  {17004} (\bibinfo {year} {2017})}\BibitemShut {NoStop}%
\bibitem [{\citenamefont {Mundy}\ \emph {et~al.}(2017)\citenamefont {Mundy},
  \citenamefont {Schaab}, \citenamefont {Kumagai}, \citenamefont {Cano},
  \citenamefont {Stengel}, \citenamefont {Krug}, \citenamefont {Gottlob},
  \citenamefont {Do{\u g}anay}, \citenamefont {Holtz}, \citenamefont {Held},
  \citenamefont {Yan}, \citenamefont {Bourret}, \citenamefont {Schneider},
  \citenamefont {Schlom}, \citenamefont {Muller}, \citenamefont {Ramesh},
  \citenamefont {Spaldin},\ and\ \citenamefont {Meier}}]{Mundy2017}%
  \BibitemOpen
  \bibfield  {author} {\bibinfo {author} {\bibfnamefont {J.~A.}\ \bibnamefont
  {Mundy}}, \bibinfo {author} {\bibfnamefont {J.}~\bibnamefont {Schaab}},
  \bibinfo {author} {\bibfnamefont {Y.}~\bibnamefont {Kumagai}}, \bibinfo
  {author} {\bibfnamefont {A.}~\bibnamefont {Cano}}, \bibinfo {author}
  {\bibfnamefont {M.}~\bibnamefont {Stengel}}, \bibinfo {author} {\bibfnamefont
  {I.~P.}\ \bibnamefont {Krug}}, \bibinfo {author} {\bibfnamefont {D.~M.}\
  \bibnamefont {Gottlob}}, \bibinfo {author} {\bibfnamefont {H.}~\bibnamefont
  {Do{\u g}anay}}, \bibinfo {author} {\bibfnamefont {M.~E.}\ \bibnamefont
  {Holtz}}, \bibinfo {author} {\bibfnamefont {R.}~\bibnamefont {Held}},
  \bibinfo {author} {\bibfnamefont {Z.}~\bibnamefont {Yan}}, \bibinfo {author}
  {\bibfnamefont {E.}~\bibnamefont {Bourret}}, \bibinfo {author} {\bibfnamefont
  {C.~M.}\ \bibnamefont {Schneider}}, \bibinfo {author} {\bibfnamefont {D.~G.}\
  \bibnamefont {Schlom}}, \bibinfo {author} {\bibfnamefont {D.~A.}\
  \bibnamefont {Muller}}, \bibinfo {author} {\bibfnamefont {R.}~\bibnamefont
  {Ramesh}}, \bibinfo {author} {\bibfnamefont {N.~A.}\ \bibnamefont {Spaldin}},
  \ and\ \bibinfo {author} {\bibfnamefont {D.}~\bibnamefont {Meier}},\
  }\href@noop {} {\bibfield  {journal} {\bibinfo  {journal} {Nature Materials}\
  }\textbf {\bibinfo {volume} {16}},\ \bibinfo {pages} {622} (\bibinfo {year}
  {2017})}\BibitemShut {NoStop}%
\bibitem [{\citenamefont {Turner}\ \emph {et~al.}(2018)\citenamefont {Turner},
  \citenamefont {McConville}, \citenamefont {McCartan}, \citenamefont
  {Campbell}, \citenamefont {Schaab}, \citenamefont {McQuaid}, \citenamefont
  {Kumar},\ and\ \citenamefont {Gregg}}]{Turner2018}%
  \BibitemOpen
  \bibfield  {author} {\bibinfo {author} {\bibfnamefont {P.~W.}\ \bibnamefont
  {Turner}}, \bibinfo {author} {\bibfnamefont {J.~P.~V.}\ \bibnamefont
  {McConville}}, \bibinfo {author} {\bibfnamefont {S.~J.}\ \bibnamefont
  {McCartan}}, \bibinfo {author} {\bibfnamefont {M.~H.}\ \bibnamefont
  {Campbell}}, \bibinfo {author} {\bibfnamefont {J.}~\bibnamefont {Schaab}},
  \bibinfo {author} {\bibfnamefont {R.~G.~P.}\ \bibnamefont {McQuaid}},
  \bibinfo {author} {\bibfnamefont {A.}~\bibnamefont {Kumar}}, \ and\ \bibinfo
  {author} {\bibfnamefont {J.~M.}\ \bibnamefont {Gregg}},\ }\href@noop {}
  {\bibfield  {journal} {\bibinfo  {journal} {Nano Letters}\ } (\bibinfo {year}
  {2018})}\BibitemShut {NoStop}%
\bibitem [{\citenamefont {Akamatsu}\ \emph {et~al.}(2018)\citenamefont
  {Akamatsu}, \citenamefont {Yuan}, \citenamefont {Stoica}, \citenamefont
  {Stone}, \citenamefont {Yang}, \citenamefont {Hong}, \citenamefont {Lei},
  \citenamefont {Zhu}, \citenamefont {Haislmaier}, \citenamefont {Freeland},
  \citenamefont {Chen}, \citenamefont {Wen},\ and\ \citenamefont
  {Gopalan}}]{Akamatsu2018}%
  \BibitemOpen
  \bibfield  {author} {\bibinfo {author} {\bibfnamefont {H.}~\bibnamefont
  {Akamatsu}}, \bibinfo {author} {\bibfnamefont {Y.}~\bibnamefont {Yuan}},
  \bibinfo {author} {\bibfnamefont {V.~A.}\ \bibnamefont {Stoica}}, \bibinfo
  {author} {\bibfnamefont {G.}~\bibnamefont {Stone}}, \bibinfo {author}
  {\bibfnamefont {T.}~\bibnamefont {Yang}}, \bibinfo {author} {\bibfnamefont
  {Z.}~\bibnamefont {Hong}}, \bibinfo {author} {\bibfnamefont {S.}~\bibnamefont
  {Lei}}, \bibinfo {author} {\bibfnamefont {Y.}~\bibnamefont {Zhu}}, \bibinfo
  {author} {\bibfnamefont {R.~C.}\ \bibnamefont {Haislmaier}}, \bibinfo
  {author} {\bibfnamefont {J.~W.}\ \bibnamefont {Freeland}}, \bibinfo {author}
  {\bibfnamefont {L.-Q.}\ \bibnamefont {Chen}}, \bibinfo {author}
  {\bibfnamefont {H.}~\bibnamefont {Wen}}, \ and\ \bibinfo {author}
  {\bibfnamefont {V.}~\bibnamefont {Gopalan}},\ }\href@noop {} {\bibfield
  {journal} {\bibinfo  {journal} {Phys Rev Lett}\ }\textbf {\bibinfo {volume}
  {120}} (\bibinfo {year} {2018})}\BibitemShut {NoStop}%
\bibitem [{\citenamefont {Schaab}\ \emph {et~al.}(2018)\citenamefont {Schaab},
  \citenamefont {Skj{\ae}rv{\o}}, \citenamefont {Krohns}, \citenamefont {Dai},
  \citenamefont {Holtz}, \citenamefont {Cano}, \citenamefont {Lilienblum},
  \citenamefont {Yan}, \citenamefont {Bourret}, \citenamefont {Muller},
  \citenamefont {Fiebig}, \citenamefont {Selbach},\ and\ \citenamefont
  {Meier}}]{Schaab2018}%
  \BibitemOpen
  \bibfield  {author} {\bibinfo {author} {\bibfnamefont {J.}~\bibnamefont
  {Schaab}}, \bibinfo {author} {\bibfnamefont {S.~H.}\ \bibnamefont
  {Skj{\ae}rv{\o}}}, \bibinfo {author} {\bibfnamefont {S.}~\bibnamefont
  {Krohns}}, \bibinfo {author} {\bibfnamefont {X.}~\bibnamefont {Dai}},
  \bibinfo {author} {\bibfnamefont {M.~E.}\ \bibnamefont {Holtz}}, \bibinfo
  {author} {\bibfnamefont {A.}~\bibnamefont {Cano}}, \bibinfo {author}
  {\bibfnamefont {M.}~\bibnamefont {Lilienblum}}, \bibinfo {author}
  {\bibfnamefont {Z.}~\bibnamefont {Yan}}, \bibinfo {author} {\bibfnamefont
  {E.}~\bibnamefont {Bourret}}, \bibinfo {author} {\bibfnamefont {D.~A.}\
  \bibnamefont {Muller}}, \bibinfo {author} {\bibfnamefont {M.}~\bibnamefont
  {Fiebig}}, \bibinfo {author} {\bibfnamefont {S.~M.}\ \bibnamefont {Selbach}},
  \ and\ \bibinfo {author} {\bibfnamefont {D.}~\bibnamefont {Meier}},\
  }\href@noop {} {\bibfield  {journal} {\bibinfo  {journal} {Nature
  Nanotechnology}\ }\textbf {\bibinfo {volume} {13}},\ \bibinfo {pages} {1028}
  (\bibinfo {year} {2018})}\BibitemShut {NoStop}%
\bibitem [{\citenamefont {Tselev}\ \emph {et~al.}(2016)\citenamefont {Tselev},
  \citenamefont {Yu}, \citenamefont {Cao}, \citenamefont {Dedon}, \citenamefont
  {Martin}, \citenamefont {Kalinin},\ and\ \citenamefont
  {Maksymovych}}]{Tselev2016}%
  \BibitemOpen
  \bibfield  {author} {\bibinfo {author} {\bibfnamefont {A.}~\bibnamefont
  {Tselev}}, \bibinfo {author} {\bibfnamefont {P.}~\bibnamefont {Yu}}, \bibinfo
  {author} {\bibfnamefont {Y.}~\bibnamefont {Cao}}, \bibinfo {author}
  {\bibfnamefont {L.~R.}\ \bibnamefont {Dedon}}, \bibinfo {author}
  {\bibfnamefont {L.~W.}\ \bibnamefont {Martin}}, \bibinfo {author}
  {\bibfnamefont {S.~V.}\ \bibnamefont {Kalinin}}, \ and\ \bibinfo {author}
  {\bibfnamefont {P.}~\bibnamefont {Maksymovych}},\ }\href@noop {} {\bibfield
  {journal} {\bibinfo  {journal} {Nature Communications}\ }\textbf {\bibinfo
  {volume} {7}},\ \bibinfo {pages} {11630} (\bibinfo {year}
  {2016})}\BibitemShut {NoStop}%
\bibitem [{\citenamefont {Prosandeev}\ \emph {et~al.}(2018)\citenamefont
  {Prosandeev}, \citenamefont {Yang}, \citenamefont {Paillard},\ and\
  \citenamefont {Bellaiche}}]{Prosandeev2018}%
  \BibitemOpen
  \bibfield  {author} {\bibinfo {author} {\bibfnamefont {S.}~\bibnamefont
  {Prosandeev}}, \bibinfo {author} {\bibfnamefont {Y.}~\bibnamefont {Yang}},
  \bibinfo {author} {\bibfnamefont {C.}~\bibnamefont {Paillard}}, \ and\
  \bibinfo {author} {\bibfnamefont {L.}~\bibnamefont {Bellaiche}},\ }\href@noop
  {} {\bibfield  {journal} {\bibinfo  {journal} {npj Comput Mater}\ }\textbf
  {\bibinfo {volume} {4}} (\bibinfo {year} {2018})}\BibitemShut {NoStop}%
\bibitem [{\citenamefont {Hlinka}\ \emph {et~al.}(2017)\citenamefont {Hlinka},
  \citenamefont {Pa{\'{s}}ciak}, \citenamefont {Körbel},\ and\ \citenamefont
  {Marton}}]{Hlinka2017}%
  \BibitemOpen
  \bibfield  {author} {\bibinfo {author} {\bibfnamefont {J.}~\bibnamefont
  {Hlinka}}, \bibinfo {author} {\bibfnamefont {M.}~\bibnamefont
  {Pa{\'{s}}ciak}}, \bibinfo {author} {\bibfnamefont {S.}~\bibnamefont
  {Körbel}}, \ and\ \bibinfo {author} {\bibfnamefont {P.}~\bibnamefont
  {Marton}},\ }\href@noop {} {\bibfield  {journal} {\bibinfo  {journal} {Phys.
  Rev. Lett.}\ }\textbf {\bibinfo {volume} {119}} (\bibinfo {year}
  {2017})}\BibitemShut {NoStop}%
\bibitem [{\citenamefont {Manipatruni}\ \emph {et~al.}(2018)\citenamefont
  {Manipatruni}, \citenamefont {Nikonov}, \citenamefont {Lin}, \citenamefont
  {Prasad}, \citenamefont {Huang}, \citenamefont {Damodaran}, \citenamefont
  {Chen}, \citenamefont {Ramesh},\ and\ \citenamefont {Young}}]{Ramesh2018}%
  \BibitemOpen
  \bibfield  {author} {\bibinfo {author} {\bibfnamefont {S.}~\bibnamefont
  {Manipatruni}}, \bibinfo {author} {\bibfnamefont {D.~E.}\ \bibnamefont
  {Nikonov}}, \bibinfo {author} {\bibfnamefont {C.-C.}\ \bibnamefont {Lin}},
  \bibinfo {author} {\bibfnamefont {B.}~\bibnamefont {Prasad}}, \bibinfo
  {author} {\bibfnamefont {Y.-L.}\ \bibnamefont {Huang}}, \bibinfo {author}
  {\bibfnamefont {A.~R.}\ \bibnamefont {Damodaran}}, \bibinfo {author}
  {\bibfnamefont {Z.}~\bibnamefont {Chen}}, \bibinfo {author} {\bibfnamefont
  {R.}~\bibnamefont {Ramesh}}, \ and\ \bibinfo {author} {\bibfnamefont {I.~A.}\
  \bibnamefont {Young}},\ }\href@noop {} {\bibfield  {journal} {\bibinfo
  {journal} {Sci. Adv.}\ }\textbf {\bibinfo {volume} {4}},\ \bibinfo {pages}
  {eaat4229} (\bibinfo {year} {2018})}\BibitemShut {NoStop}%
\bibitem [{\citenamefont {Marton}\ \emph {et~al.}(2010)\citenamefont {Marton},
  \citenamefont {Rychetsky},\ and\ \citenamefont {Hlinka}}]{Marton2010}%
  \BibitemOpen
  \bibfield  {author} {\bibinfo {author} {\bibfnamefont {P.}~\bibnamefont
  {Marton}}, \bibinfo {author} {\bibfnamefont {I.}~\bibnamefont {Rychetsky}}, \
  and\ \bibinfo {author} {\bibfnamefont {J.}~\bibnamefont {Hlinka}},\
  }\href@noop {} {\bibfield  {journal} {\bibinfo  {journal} {Phys. Rev. B}\
  }\textbf {\bibinfo {volume} {81}} (\bibinfo {year} {2010})}\BibitemShut
  {NoStop}%
\bibitem [{\citenamefont {Ziegler}\ \emph {et~al.}(2013)\citenamefont
  {Ziegler}, \citenamefont {Martens}, \citenamefont {Giamarchi},\ and\
  \citenamefont {Paruch}}]{Ziegler2013}%
  \BibitemOpen
  \bibfield  {author} {\bibinfo {author} {\bibfnamefont {B.}~\bibnamefont
  {Ziegler}}, \bibinfo {author} {\bibfnamefont {K.}~\bibnamefont {Martens}},
  \bibinfo {author} {\bibfnamefont {T.}~\bibnamefont {Giamarchi}}, \ and\
  \bibinfo {author} {\bibfnamefont {P.}~\bibnamefont {Paruch}},\ }\href@noop {}
  {\bibfield  {journal} {\bibinfo  {journal} {Phys. Rev. Lett.}\ }\textbf
  {\bibinfo {volume} {111}},\ \bibinfo {pages} {247604} (\bibinfo {year}
  {2013})}\BibitemShut {NoStop}%
\bibitem [{\citenamefont {Domingo}\ \emph {et~al.}(2017)\citenamefont
  {Domingo}, \citenamefont {Farokhipoor}, \citenamefont {Santiso},
  \citenamefont {Noheda},\ and\ \citenamefont {Catalan}}]{Domingo2017}%
  \BibitemOpen
  \bibfield  {author} {\bibinfo {author} {\bibfnamefont {N.}~\bibnamefont
  {Domingo}}, \bibinfo {author} {\bibfnamefont {S.}~\bibnamefont
  {Farokhipoor}}, \bibinfo {author} {\bibfnamefont {J.}~\bibnamefont
  {Santiso}}, \bibinfo {author} {\bibfnamefont {B.}~\bibnamefont {Noheda}}, \
  and\ \bibinfo {author} {\bibfnamefont {G.}~\bibnamefont {Catalan}},\
  }\href@noop {} {\bibfield  {journal} {\bibinfo  {journal} {Journal of
  Physics: Condensed Matter}\ }\textbf {\bibinfo {volume} {29}},\ \bibinfo
  {pages} {334003} (\bibinfo {year} {2017})}\BibitemShut {NoStop}%
\bibitem [{\citenamefont {Royo}\ \emph {et~al.}(2017)\citenamefont {Royo},
  \citenamefont {Escorihuela-Sayalero}, \citenamefont {\'I\~niguez},\ and\
  \citenamefont {Rurali}}]{Royo2017}%
  \BibitemOpen
  \bibfield  {author} {\bibinfo {author} {\bibfnamefont {M.}~\bibnamefont
  {Royo}}, \bibinfo {author} {\bibfnamefont {C.}~\bibnamefont
  {Escorihuela-Sayalero}}, \bibinfo {author} {\bibfnamefont {J.}~\bibnamefont
  {\'I\~niguez}}, \ and\ \bibinfo {author} {\bibfnamefont {R.}~\bibnamefont
  {Rurali}},\ }\href@noop {} {\bibfield  {journal} {\bibinfo  {journal} {Phys.
  Rev. Materials}\ }\textbf {\bibinfo {volume} {1}},\ \bibinfo {pages} {051402}
  (\bibinfo {year} {2017})}\BibitemShut {NoStop}%
\bibitem [{\citenamefont {Stefani}\ \emph {et~al.}()\citenamefont {Stefani},
  \citenamefont {Ponet}, \citenamefont {Shapovalov}, \citenamefont {Chen},
  \citenamefont {Stengel}, \citenamefont {Artyukhin}, \citenamefont {Catalan},\
  and\ \citenamefont {Domingo}}]{Stefani2025}%
  \BibitemOpen
  \bibfield  {author} {\bibinfo {author} {\bibfnamefont {C.}~\bibnamefont
  {Stefani}}, \bibinfo {author} {\bibfnamefont {L.}~\bibnamefont {Ponet}},
  \bibinfo {author} {\bibfnamefont {K.}~\bibnamefont {Shapovalov}}, \bibinfo
  {author} {\bibfnamefont {P.}~\bibnamefont {Chen}}, \bibinfo {author}
  {\bibfnamefont {M.}~\bibnamefont {Stengel}}, \bibinfo {author} {\bibfnamefont
  {S.}~\bibnamefont {Artyukhin}}, \bibinfo {author} {\bibfnamefont
  {G.}~\bibnamefont {Catalan}}, \ and\ \bibinfo {author} {\bibfnamefont
  {N.}~\bibnamefont {Domingo}},\ }\href@noop {} {\bibinfo  {journal} {in
  preparation}\ }\BibitemShut {NoStop}%
\bibitem [{\citenamefont {Artyukhin}\ \emph {et~al.}(2013)\citenamefont
  {Artyukhin}, \citenamefont {Delaney}, \citenamefont {Spaldin},\ and\
  \citenamefont {Mostovoy}}]{Artyukhin2013}%
  \BibitemOpen
\bibfield  {journal} {  }\bibfield  {author} {\bibinfo {author} {\bibfnamefont
  {S.}~\bibnamefont {Artyukhin}}, \bibinfo {author} {\bibfnamefont {K.~T.}\
  \bibnamefont {Delaney}}, \bibinfo {author} {\bibfnamefont {N.~A.}\
  \bibnamefont {Spaldin}}, \ and\ \bibinfo {author} {\bibfnamefont
  {M.}~\bibnamefont {Mostovoy}},\ }\href@noop {} {\bibfield  {journal}
  {\bibinfo  {journal} {Nature Materials}\ }\textbf {\bibinfo {volume} {13}},\
  \bibinfo {pages} {42} (\bibinfo {year} {2013})}\BibitemShut {NoStop}%
\bibitem [{\citenamefont {Wang}\ \emph {et~al.}(2014)\citenamefont {Wang},
  \citenamefont {Mostovoy}, \citenamefont {Han}, \citenamefont {Horibe},
  \citenamefont {Aoki}, \citenamefont {Zhu},\ and\ \citenamefont
  {Cheong}}]{Wang2014}%
  \BibitemOpen
  \bibfield  {author} {\bibinfo {author} {\bibfnamefont {X.}~\bibnamefont
  {Wang}}, \bibinfo {author} {\bibfnamefont {M.}~\bibnamefont {Mostovoy}},
  \bibinfo {author} {\bibfnamefont {M.~G.}\ \bibnamefont {Han}}, \bibinfo
  {author} {\bibfnamefont {Y.}~\bibnamefont {Horibe}}, \bibinfo {author}
  {\bibfnamefont {T.}~\bibnamefont {Aoki}}, \bibinfo {author} {\bibfnamefont
  {Y.}~\bibnamefont {Zhu}}, \ and\ \bibinfo {author} {\bibfnamefont {S.-W.}\
  \bibnamefont {Cheong}},\ }\href@noop {} {\bibfield  {journal} {\bibinfo
  {journal} {Phys. Rev. Lett.}\ }\textbf {\bibinfo {volume} {112}},\ \bibinfo
  {pages} {247601} (\bibinfo {year} {2014})}\BibitemShut {NoStop}%
\bibitem [{\citenamefont {Janovec}(1976)}]{Janovec1976}%
  \BibitemOpen
  \bibfield  {author} {\bibinfo {author} {\bibfnamefont {V.}~\bibnamefont
  {Janovec}},\ }\href@noop {} {\bibfield  {journal} {\bibinfo  {journal}
  {Ferroelectrics}\ }\textbf {\bibinfo {volume} {12}},\ \bibinfo {pages} {43}
  (\bibinfo {year} {1976})}\BibitemShut {NoStop}%
\bibitem [{\citenamefont {Fousek}\ and\ \citenamefont
  {Janovec}(1969)}]{Fousek1969}%
  \BibitemOpen
  \bibfield  {author} {\bibinfo {author} {\bibfnamefont {J.}~\bibnamefont
  {Fousek}}\ and\ \bibinfo {author} {\bibfnamefont {V.}~\bibnamefont
  {Janovec}},\ }\href@noop {} {\bibfield  {journal} {\bibinfo  {journal} {J
  Appl Phys}\ }\textbf {\bibinfo {volume} {40}},\ \bibinfo {pages} {135}
  (\bibinfo {year} {1969})}\BibitemShut {NoStop}%
\bibitem [{\citenamefont {Ishibashi}\ and\ \citenamefont
  {Salje}(2002)}]{Ishibashi2002}%
  \BibitemOpen
  \bibfield  {author} {\bibinfo {author} {\bibfnamefont {Y.}~\bibnamefont
  {Ishibashi}}\ and\ \bibinfo {author} {\bibfnamefont {E.}~\bibnamefont
  {Salje}},\ }\href@noop {} {\bibfield  {journal} {\bibinfo  {journal} {Journal
  of the Physical Society of Japan}\ }\textbf {\bibinfo {volume} {71}},\
  \bibinfo {pages} {2800} (\bibinfo {year} {2002})}\BibitemShut {NoStop}%
\bibitem [{\citenamefont {Ishibashi}\ \emph {et~al.}(2005)\citenamefont
  {Ishibashi}, \citenamefont {Iwata},\ and\ \citenamefont
  {Salje}}]{Yoshihiro2005}%
  \BibitemOpen
  \bibfield  {author} {\bibinfo {author} {\bibfnamefont {Y.}~\bibnamefont
  {Ishibashi}}, \bibinfo {author} {\bibfnamefont {M.}~\bibnamefont {Iwata}}, \
  and\ \bibinfo {author} {\bibfnamefont {E.}~\bibnamefont {Salje}},\
  }\href@noop {} {\bibfield  {journal} {\bibinfo  {journal} {Japanese Journal
  of Applied Physics}\ }\textbf {\bibinfo {volume} {44}},\ \bibinfo {pages}
  {7512} (\bibinfo {year} {2005})}\BibitemShut {NoStop}%
\bibitem [{\citenamefont {Conti}\ and\ \citenamefont
  {Weikard}(2004)}]{Conti2004}%
  \BibitemOpen
  \bibfield  {author} {\bibinfo {author} {\bibfnamefont {S.}~\bibnamefont
  {Conti}}\ and\ \bibinfo {author} {\bibfnamefont {U.}~\bibnamefont
  {Weikard}},\ }\href@noop {} {\bibfield  {journal} {\bibinfo  {journal} {The
  European Physical Journal B - Condensed Matter and Complex Systems}\ }\textbf
  {\bibinfo {volume} {41}},\ \bibinfo {pages} {413} (\bibinfo {year}
  {2004})}\BibitemShut {NoStop}%
\bibitem [{\citenamefont {Li}\ \emph {et~al.}(2018)\citenamefont {Li},
  \citenamefont {Jokisaari}, \citenamefont {Zhang}, \citenamefont {Cheng},
  \citenamefont {Yan}, \citenamefont {Heikes}, \citenamefont {Lin},
  \citenamefont {Gadre}, \citenamefont {Schlom}, \citenamefont {Chen},\ and\
  \citenamefont {Pan}}]{Linze2018}%
  \BibitemOpen
  \bibfield  {author} {\bibinfo {author} {\bibfnamefont {L.}~\bibnamefont
  {Li}}, \bibinfo {author} {\bibfnamefont {J.~R.}\ \bibnamefont {Jokisaari}},
  \bibinfo {author} {\bibfnamefont {Y.}~\bibnamefont {Zhang}}, \bibinfo
  {author} {\bibfnamefont {X.}~\bibnamefont {Cheng}}, \bibinfo {author}
  {\bibfnamefont {X.}~\bibnamefont {Yan}}, \bibinfo {author} {\bibfnamefont
  {C.}~\bibnamefont {Heikes}}, \bibinfo {author} {\bibfnamefont
  {Q.}~\bibnamefont {Lin}}, \bibinfo {author} {\bibfnamefont {C.}~\bibnamefont
  {Gadre}}, \bibinfo {author} {\bibfnamefont {D.~G.}\ \bibnamefont {Schlom}},
  \bibinfo {author} {\bibfnamefont {L.-Q.}\ \bibnamefont {Chen}}, \ and\
  \bibinfo {author} {\bibfnamefont {X.}~\bibnamefont {Pan}},\ }\href@noop {}
  {\bibfield  {journal} {\bibinfo  {journal} {Advanced Materials}\ }\textbf
  {\bibinfo {volume} {30}},\ \bibinfo {pages} {1802737} (\bibinfo {year}
  {2018})}\BibitemShut {NoStop}%
\bibitem [{\citenamefont {{Salje}}(1993)}]{Salje1993}%
  \BibitemOpen
  \bibfield  {author} {\bibinfo {author} {\bibfnamefont {E.~K.}\ \bibnamefont
  {{Salje}}},\ }\href@noop {} {\emph {\bibinfo {title} {Phase Transitions in
  Ferroelastic and Co-elastic Crystals}}}\ (\bibinfo  {publisher} {Cambridge,
  UK: Cambridge University Press},\ \bibinfo {year} {1993})\ p.\ \bibinfo
  {pages} {296}\BibitemShut {NoStop}%
\bibitem [{\citenamefont {Salje}\ \emph {et~al.}(2016)\citenamefont {Salje},
  \citenamefont {Aktas},\ and\ \citenamefont {Ding}}]{Salje2016}%
  \BibitemOpen
  \bibfield  {author} {\bibinfo {author} {\bibfnamefont {E.~K.~H.}\
  \bibnamefont {Salje}}, \bibinfo {author} {\bibfnamefont {O.}~\bibnamefont
  {Aktas}}, \ and\ \bibinfo {author} {\bibfnamefont {X.}~\bibnamefont {Ding}},\
  }\enquote {\bibinfo {title} {Functional topologies in (multi-) ferroics: The
  ferroelastic template},}\ in\ \href@noop {} {\emph {\bibinfo {booktitle}
  {Topological Structures in Ferroic Materials: Domain Walls, Vortices and
  Skyrmions}}},\ \bibinfo {editor} {edited by\ \bibinfo {editor} {\bibfnamefont
  {J.}~\bibnamefont {Seidel}}}\ (\bibinfo  {publisher} {Springer International
  Publishing},\ \bibinfo {address} {Cham},\ \bibinfo {year} {2016})\ pp.\
  \bibinfo {pages} {83--101}\BibitemShut {NoStop}%
\bibitem [{\citenamefont {Griffin}\ \emph {et~al.}(2012)\citenamefont
  {Griffin}, \citenamefont {Lilienblum}, \citenamefont {Delaney}, \citenamefont
  {Kumagai}, \citenamefont {Fiebig},\ and\ \citenamefont
  {Spaldin}}]{Griffin12}%
  \BibitemOpen
  \bibfield  {author} {\bibinfo {author} {\bibfnamefont {S.~M.}\ \bibnamefont
  {Griffin}}, \bibinfo {author} {\bibfnamefont {M.}~\bibnamefont {Lilienblum}},
  \bibinfo {author} {\bibfnamefont {K.~T.}\ \bibnamefont {Delaney}}, \bibinfo
  {author} {\bibfnamefont {Y.}~\bibnamefont {Kumagai}}, \bibinfo {author}
  {\bibfnamefont {M.}~\bibnamefont {Fiebig}}, \ and\ \bibinfo {author}
  {\bibfnamefont {N.~A.}\ \bibnamefont {Spaldin}},\ }\href@noop {} {\bibfield
  {journal} {\bibinfo  {journal} {Phys. Rev. X}\ }\textbf {\bibinfo {volume}
  {2}},\ \bibinfo {pages} {041022} (\bibinfo {year} {2012})}\BibitemShut
  {NoStop}%
\bibitem [{\citenamefont {Chae}\ \emph {et~al.}(2012)\citenamefont {Chae},
  \citenamefont {Lee}, \citenamefont {Horibe}, \citenamefont {Tanimura},
  \citenamefont {Mori}, \citenamefont {Gao}, \citenamefont {Carr},\ and\
  \citenamefont {Cheong}}]{Chae2012}%
  \BibitemOpen
  \bibfield  {author} {\bibinfo {author} {\bibfnamefont {S.~C.}\ \bibnamefont
  {Chae}}, \bibinfo {author} {\bibfnamefont {N.}~\bibnamefont {Lee}}, \bibinfo
  {author} {\bibfnamefont {Y.}~\bibnamefont {Horibe}}, \bibinfo {author}
  {\bibfnamefont {M.}~\bibnamefont {Tanimura}}, \bibinfo {author}
  {\bibfnamefont {S.}~\bibnamefont {Mori}}, \bibinfo {author} {\bibfnamefont
  {B.}~\bibnamefont {Gao}}, \bibinfo {author} {\bibfnamefont {S.}~\bibnamefont
  {Carr}}, \ and\ \bibinfo {author} {\bibfnamefont {S.-W.}\ \bibnamefont
  {Cheong}},\ }\href@noop {} {\bibfield  {journal} {\bibinfo  {journal} {Phys.
  Rev. Lett.}\ }\textbf {\bibinfo {volume} {108}},\ \bibinfo {pages} {167603}
  (\bibinfo {year} {2012})}\BibitemShut {NoStop}%
\bibitem [{\citenamefont {Nambu}\ and\ \citenamefont
  {Sagala}(1994)}]{Nambu1994}%
  \BibitemOpen
  \bibfield  {author} {\bibinfo {author} {\bibfnamefont {S.}~\bibnamefont
  {Nambu}}\ and\ \bibinfo {author} {\bibfnamefont {D.~A.}\ \bibnamefont
  {Sagala}},\ }\href@noop {} {\bibfield  {journal} {\bibinfo  {journal} {Phys.
  Rev. B}\ }\textbf {\bibinfo {volume} {50}},\ \bibinfo {pages} {5838}
  (\bibinfo {year} {1994})}\BibitemShut {NoStop}%
\bibitem [{\citenamefont {Park}\ \emph {et~al.}(2018)\citenamefont {Park},
  \citenamefont {Wang}, \citenamefont {Das}, \citenamefont {Chae},
  \citenamefont {Chung}, \citenamefont {Yoon}, \citenamefont {Chen},
  \citenamefont {Yang},\ and\ \citenamefont {Noh}}]{Park2018}%
  \BibitemOpen
  \bibfield  {author} {\bibinfo {author} {\bibfnamefont {S.~M.}\ \bibnamefont
  {Park}}, \bibinfo {author} {\bibfnamefont {B.}~\bibnamefont {Wang}}, \bibinfo
  {author} {\bibfnamefont {S.}~\bibnamefont {Das}}, \bibinfo {author}
  {\bibfnamefont {S.~C.}\ \bibnamefont {Chae}}, \bibinfo {author}
  {\bibfnamefont {J.-S.}\ \bibnamefont {Chung}}, \bibinfo {author}
  {\bibfnamefont {J.-G.}\ \bibnamefont {Yoon}}, \bibinfo {author}
  {\bibfnamefont {L.-Q.}\ \bibnamefont {Chen}}, \bibinfo {author}
  {\bibfnamefont {S.~M.}\ \bibnamefont {Yang}}, \ and\ \bibinfo {author}
  {\bibfnamefont {T.~W.}\ \bibnamefont {Noh}},\ }\href@noop {} {\bibfield
  {journal} {\bibinfo  {journal} {Nature Nanotechnology}\ }\textbf {\bibinfo
  {volume} {13}},\ \bibinfo {pages} {366} (\bibinfo {year} {2018})}\BibitemShut
  {NoStop}%
\bibitem [{\citenamefont {Saj~Mohan}\ \emph {et~al.}(2019)\citenamefont
  {Saj~Mohan}, \citenamefont {Bandyopadhyay}, \citenamefont {Jogi},
  \citenamefont {Bhattacharya},\ and\ \citenamefont
  {Ramadurai}}]{SajMohan2019}%
  \BibitemOpen
  \bibfield  {author} {\bibinfo {author} {\bibfnamefont {M.~M.}\ \bibnamefont
  {Saj~Mohan}}, \bibinfo {author} {\bibfnamefont {S.}~\bibnamefont
  {Bandyopadhyay}}, \bibinfo {author} {\bibfnamefont {T.}~\bibnamefont {Jogi}},
  \bibinfo {author} {\bibfnamefont {S.}~\bibnamefont {Bhattacharya}}, \ and\
  \bibinfo {author} {\bibfnamefont {R.}~\bibnamefont {Ramadurai}},\ }\href@noop
  {} {\bibfield  {journal} {\bibinfo  {journal} {Journal of Applied Physics}\
  }\textbf {\bibinfo {volume} {125}},\ \bibinfo {pages} {012501} (\bibinfo
  {year} {2019})}\BibitemShut {NoStop}%
\bibitem [{\citenamefont {Aln{\ae}s}\ \emph {et~al.}(2015)\citenamefont
  {Aln{\ae}s}, \citenamefont {Blechta}, \citenamefont {Hake}, \citenamefont
  {Johansson}, \citenamefont {Kehlet}, \citenamefont {Logg}, \citenamefont
  {Richardson}, \citenamefont {Ring}, \citenamefont {Rognes},\ and\
  \citenamefont {Wells}}]{AlnaesBlechta2015a}%
  \BibitemOpen
  \bibfield  {author} {\bibinfo {author} {\bibfnamefont {M.~S.}\ \bibnamefont
  {Aln{\ae}s}}, \bibinfo {author} {\bibfnamefont {J.}~\bibnamefont {Blechta}},
  \bibinfo {author} {\bibfnamefont {J.}~\bibnamefont {Hake}}, \bibinfo {author}
  {\bibfnamefont {A.}~\bibnamefont {Johansson}}, \bibinfo {author}
  {\bibfnamefont {B.}~\bibnamefont {Kehlet}}, \bibinfo {author} {\bibfnamefont
  {A.}~\bibnamefont {Logg}}, \bibinfo {author} {\bibfnamefont {C.}~\bibnamefont
  {Richardson}}, \bibinfo {author} {\bibfnamefont {J.}~\bibnamefont {Ring}},
  \bibinfo {author} {\bibfnamefont {M.~E.}\ \bibnamefont {Rognes}}, \ and\
  \bibinfo {author} {\bibfnamefont {G.~N.}\ \bibnamefont {Wells}},\ }\href@noop
  {} {\bibfield  {journal} {\bibinfo  {journal} {Archive of Numerical
  Software}\ }\textbf {\bibinfo {volume} {3}} (\bibinfo {year}
  {2015})}\BibitemShut {NoStop}%
\bibitem [{\citenamefont {Logg}\ \emph {et~al.}(2012)\citenamefont {Logg},
  \citenamefont {Mardal}, \citenamefont {Wells} \emph
  {et~al.}}]{LoggMardalEtAl2012a}%
  \BibitemOpen
  \bibfield  {author} {\bibinfo {author} {\bibfnamefont {A.}~\bibnamefont
  {Logg}}, \bibinfo {author} {\bibfnamefont {K.-A.}\ \bibnamefont {Mardal}},
  \bibinfo {author} {\bibfnamefont {G.~N.}\ \bibnamefont {Wells}},  \emph
  {et~al.},\ }\href@noop {} {\emph {\bibinfo {title} {Automated Solution of
  Differential Equations by the Finite Element Method}}}\ (\bibinfo
  {publisher} {Springer},\ \bibinfo {year} {2012})\BibitemShut {NoStop}%
\bibitem [{\citenamefont {Marton}(2018)}]{Marton2018}%
  \BibitemOpen
  \bibfield  {author} {\bibinfo {author} {\bibfnamefont {P.}~\bibnamefont
  {Marton}},\ }\href@noop {} {\bibfield  {journal} {\bibinfo  {journal} {Phase
  Transitions}\ }\textbf {\bibinfo {volume} {91}},\ \bibinfo {pages} {959}
  (\bibinfo {year} {2018})}\BibitemShut {NoStop}%
\bibitem [{\citenamefont {Sidorkin}(2012)}]{Sidorkin2012}%
  \BibitemOpen
  \bibfield  {author} {\bibinfo {author} {\bibfnamefont {A.~S.}\ \bibnamefont
  {Sidorkin}},\ }\href@noop {} {\bibfield  {journal} {\bibinfo  {journal}
  {Journal of Advanced Dielectrics}\ }\textbf {\bibinfo {volume} {02}},\
  \bibinfo {pages} {1230013} (\bibinfo {year} {2012})}\BibitemShut {NoStop}%
\bibitem [{\citenamefont {Wang}\ \emph {et~al.}(2011)\citenamefont {Wang},
  \citenamefont {Saal}, \citenamefont {Wu}, \citenamefont {Wang}, \citenamefont
  {Shang}, \citenamefont {Liu},\ and\ \citenamefont {Chen}}]{Wang2011}%
  \BibitemOpen
  \bibfield  {author} {\bibinfo {author} {\bibfnamefont {Y.}~\bibnamefont
  {Wang}}, \bibinfo {author} {\bibfnamefont {J.~E.}\ \bibnamefont {Saal}},
  \bibinfo {author} {\bibfnamefont {P.}~\bibnamefont {Wu}}, \bibinfo {author}
  {\bibfnamefont {J.}~\bibnamefont {Wang}}, \bibinfo {author} {\bibfnamefont
  {S.}~\bibnamefont {Shang}}, \bibinfo {author} {\bibfnamefont {Z.-K.}\
  \bibnamefont {Liu}}, \ and\ \bibinfo {author} {\bibfnamefont {L.-Q.}\
  \bibnamefont {Chen}},\ }\href@noop {} {\bibfield  {journal} {\bibinfo
  {journal} {Acta Materialia}\ }\textbf {\bibinfo {volume} {59}},\ \bibinfo
  {pages} {4229 } (\bibinfo {year} {2011})}\BibitemShut {NoStop}%
\bibitem [{\citenamefont {Landau}(2004)}]{Landau2004}%
  \BibitemOpen
  \bibfield  {author} {\bibinfo {author} {\bibfnamefont {L.~D.}\ \bibnamefont
  {Landau}},\ }\href@noop {} {\emph {\bibinfo {title} {Theory of Elasticity
  7}}}\ (\bibinfo  {publisher} {Elsevier LTD, Oxford},\ \bibinfo {year}
  {2004})\BibitemShut {NoStop}%
\end{thebibliography}%
{\bf Acknowledgements}

The authors thank X. Cheng, L.Q. Chen, Y.L. Huang, L. Zheng for stimulating discussions. We acknowledge the CINECA award under the ISCRA initiative, for the availability of high performance computing resources and support. K.L. was supported by US National Science Foundations Award No. DMR-1707372.

{\bf Author contributions}

P.C. has performed the numerical simulations. P.C. and L.P. implemented and carried out the finite-element calculations on the Landau model and wrote the initial draft of the manuscript, which was finalized by S.A and R.C. with input from all authors. S.A. conceived the project and planned the study.

{\bf Competing interests}

The authors declare no competing interests.

{\bf Methods}

In order to elucidate the essential physics that is responsible for the activation of the tilted R71 walls, we use a simplified GLD model \cite{Nambu1994}. To this end, we focus on the polar mode that connects the parent centrosymmetric phase with the ferroelectric one and consider its interactions with strains (through electrostriction), while neglecting octahedral rotations.

The free energy density, expanded near the centrosymmetric paraelectric parent structure is written as:
\begin{eqnarray}\label{eq:energy}
&&f = f_{L}+f_{G}+f_{c}+f_{q}+f_{fl},\\
&&f_{L} = \alpha_{i}P_{i}^2 + \frac{1}{2}\alpha_{ij}P_{i}^2P_{j}^2 + \alpha_{ijk}P_{i}^2P_{j}^2P_{k}^2,\\
&&f_{G} = \frac{1}{2}G_{ijkl}\partial_jP_i\partial_lP_k,\\
&&f_{c} = \frac{1}{2}\epsilon_{ij}C_{ijkl}\epsilon_{kl},\\
&&f_{q}=-\frac{1}{2}\epsilon_{ij}q_{ijkl}P_{k}P_{l},\label{eq:qpp}\\
&&f_{fl}=\frac{1}{2}f_{ijkl}(\epsilon{ij}\partial_lP_{k}-\partial_l\epsilon_{ij}P_{k})
\end{eqnarray}
where $P_i$ stands for the components of the ferroelectric polarization; the strain tensor $\epsilon_{ij}$ is related to symmetrized gradients of deformations $u_i$ as $\epsilon_{ij}=(\partial_ju_i+\partial_iu_j)/2$, where the deformation vector $\vec{u}(\vec r)$ relates points $\vec r$ in the reference structure to $\vec{r} +\vec u(\vec r)$ in the deformed structure; summation over repeated indices is implied. $f_L$ represents the distorted Mexican hat-shaped potential that determines the amplitude and anisotropy of the polarization. $f_G$ describes the energy penalty due to spatial variations of the polarization. $f_c$ describes elastic energy, while $f_q$ is the electrostriction term that refers to the interaction between polarization and strain. The flexoelectric coupling $f_f$, that describes interactions of strain gradient with the polarization, was also included as in Ref.~\cite{Park2018}, but does not change the qualitative results reported here. 
The parameters of the model were adopted from Ref.~\cite{SajMohan2019}. 
In order to minimize the free energy Eq.~\ref{eq:energy} we solved the Euler-Lagrange equations with respect to $P_i, u_i$ using finite element method as implemented in Fenics software \cite{AlnaesBlechta2015a,LoggMardalEtAl2012a}. The real space was discretized with element dimensions 0.4~nm giving rise to a periodic lattice potential acting on the DW, that gaps the sliding mode at 10~GHz \cite{Marton2018}. This frequency an order of magnitude higher than characteristic frequencies of DW vibrations due to electrostatic effects \cite{Sidorkin2012}. The simulated thin film was 60~nm thick and 180~nm wide, and a single DW was placed in the middle. Test calculations were performed for the 360~nm wide film to validate the long-range strain profile. Zero external stress boundary conditions were applied on the top surface while $u_3=0$ was used at the bottom one. At the two ends in the $x_1$ direction, both open and twisted boundary conditions were tested to ensure the absence of boundary effects within the domain wall area. Periodic boundary conditions were applied in the translational direction ($x_2$) to mimic an infinite system.

Phonons are computed using 20*1*20 mesh with a single domain wall in the middle. The energy was minimized and the force constants were computed. The mass matrix was adjusted to fit the gap between acoustic and optical bands of the spectrum to DFT calculations \cite{Wang2011}.
\setcounter{table}{0}
\renewcommand{\thetable}{S\arabic{table}}%
\setcounter{figure}{0}
\renewcommand{\thefigure}{S\arabic{figure}}%
\setcounter{equation}{0}
\renewcommand{\theequation}{S\arabic{equation}}%

\clearpage
\onecolumngrid
\section{Supporting Information} 
The two polar phases which define neighboring domains  have to be mechanically connected by a DW plane described by the normal vector $\vec{n}$. 
The spontaneous polarizations and deformations in the two domains are denoted as $\{P_{i}',\epsilon_{ij}'\}$ and $\{P_{i}'',\epsilon_{ij}''\}$ respectively, where $|P_{i}'|=|P_{i}''|$. 
Let $\vec{d}$ be a vector in the DW plane within the reference structure.
In a polar domain the electrostriction leads to strain, therefore $\vec d$ at the boundary of the first domain goes into \cite{Landau2004}
\begin{equation}
\vec{d'}= \vec{d}+\vec\omega'\times \vec{d}+\hat\epsilon'\vec{d},
\label{eq:deformation}
\end{equation}
where the second term on the right-hand side represents a rigid rotation by angle $|\vec{\omega}'|$ around vector $\vec{\omega}'$. The third term corresponds to a deformation vector due to the strain tensor $\hat \epsilon'$. An analogous formula holds for $\vec d''$ in the second domain. To avoid dislocations and cracks between the two domains, these deformed vectors $\vec d',\vec d''$ must match at the DW, $\vec{d}'=\vec{d}''$. Therefore
\begin{equation}
\vec d'-\vec d''= (\vec\omega'-\vec\omega'')\times \vec{d}+(\hat\epsilon'-\hat\epsilon'')\vec{d}=0.
\label{eq:compat1}
\end{equation}
Taking the scalar product with $\vec d$ eliminates the vector product term leading to:
\begin{equation}
d_i(\epsilon_{ij}'-\epsilon_{ij}'')d_j=0
\label{eq:dw}
\end{equation}
Therefore the strain compatibility requires the strain components in the DW plane to be equal in the two domains across the wall.

\begin{figure*}[h]
\includegraphics[width=0.8\linewidth]{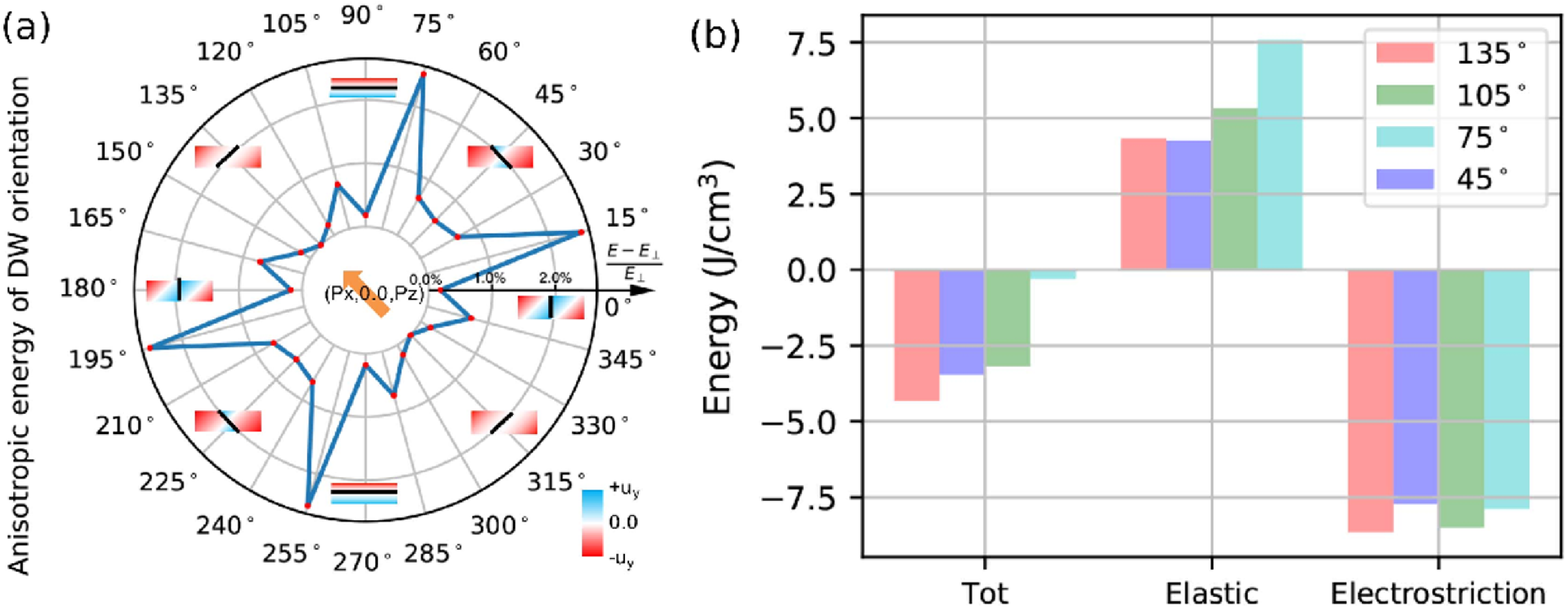}
\caption{\label{fig:aniso}(a) The dependence of energy on the orientation of the DW (shown with black lines inside a rectangular sample). The energy minimum corresponds to the wall orientation, satisfying the strain compatibility conditions. The $x_2$ component of the deformation vector inside the sample is encoded by the color. A clear crack deformation, red on one side and blue on the other, would result in a large elastic energy cost. (b) The decomposition of energy change into elastic and electrostriction contributions for some wall orientations.}
 \end{figure*}

In the bulk of each domain, the strain tensor $\epsilon_{ij}$ can be obtained from the stress free condition $\sigma_{ij} = \partial f/\partial \epsilon_{ij}$,
\begin{equation}
\epsilon_{ij} = \frac{1}{2}Q_{ijkl}P_{k}P_{l},
\label{eq:QPP}
\end{equation}
where $Q_{ijkl} = \frac{1}{2} (C^{-1})_{ijmn} \, q_{mnkl}$.
We substitute Eq.~\ref{eq:QPP} into Eq.~\ref{eq:dw}, and take into account that only $P_{2}$ changes sign across the R71 DW, to obtain $P_2d_2(P_1d_1+P_3d_3)=0$. The two solutions are $d_2=0$ and $[P_1,P_3]\perp[d_1,d_3]$, the latter meaning that all the vectors in the wall plane are perpendicular to $[P_1,0,P_3]$. 
However, in thin films the violation of the constraint does not lead to infinite elastic energy, and allows the strain to relax at the surface. Therefore, the DW may deviate from the lowest energy orientation in the bulk. 
To confirm this, we fixed $P_1,P_3$ and imposed domain walls in $P_2$ with different orientations $P_2\sim\tanh\frac{\vec n\cdot\vec r}{\lambda}$, as shown by black lines in the colored slabs in Fig.~\ref{fig:aniso}(a). As seen from Fig.~\ref{fig:aniso}(b), the DW energy is minimized when the DW normal $\vec{n}$ is parallel to $[P_1, 0, P_3]$.
Note that in the bulk the deviations from the optimal DW orientation are penalized with elastic energy scaling linearly with the DW area.
Fig.~\ref{fig:aniso}(b) shows that the energy penalty indeed comes from the elastic contribution due to strain mismatch between the domains.
\begin{figure}
     \centering
     \includegraphics[width=.5\linewidth]{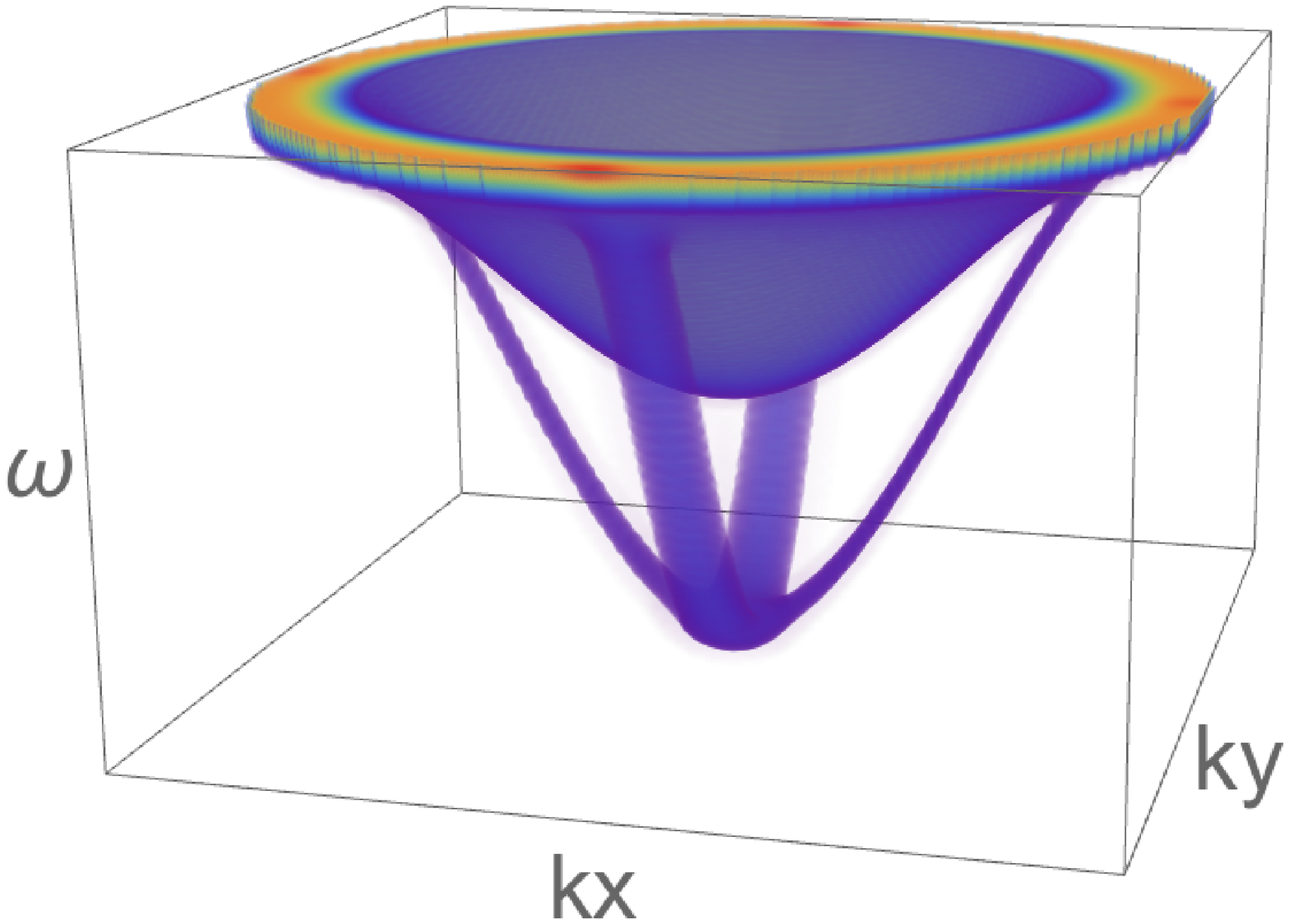}
     \caption{\label{fig:randomWalls}Schematic representation of low-energy DW-localized modes for a system with two orthogonal DW orientations. The intensities of DW branches are proportional to density of the DWs with corresponding orientations.}
 \end{figure}
\end{document}